\documentclass[12 pt]{article}

\usepackage{mathrsfs}
\usepackage[T1]{fontenc}
\usepackage{mathpazo}
\usepackage{setspace}
\usepackage{amsfonts}
\usepackage{amssymb}
\usepackage{epsfig}
\usepackage{latexsym}
\usepackage{color}
\usepackage{graphicx}
\usepackage{nicefrac}
\usepackage[latin1]{inputenc}
\usepackage{pstricks}
\usepackage{slashed}
\usepackage{multirow}

\def\hybrid{\topmargin -20pt    \oddsidemargin 0pt
        \headheight 0pt \headsep 0pt
        \textwidth 6.25in       % A4 paper
        \textheight 9.25in       % A4 paper
        \marginparwidth .875in
        \parskip 5pt plus 1pt   \jot = 1.5ex}

\hybrid

\def\baselinestretch{1.2}

\catcode`\@=11

\def\marginnote#1{}
\newcount\hour
\newcount\minute
\newtoks\amorpm
\hour=\time\divide\hour by60
\minute=\time{\multiply\hour by60 \global\advance\minute by-\hour}
\edef\standardtime{{\ifnum\hour<12 \global\amorpm={am}%
        \else\global\amorpm={pm}\advance\hour by-12 \fi
        \ifnum\hour=0 \hour=12 \fi
        \number\hour:\ifnum\minute<10 0\fi\number\minute\the\amorpm}}
\edef\militarytime{\number\hour:\ifnum\minute<10 0\fi\number\minute}
\def\draftlabel#1{{\@bsphack\if@filesw {\let\thepage\relax
   \xdef\@gtempa{\write\@auxout{\string
      \newlabel{#1}{{\@currentlabel}{\thepage}}}}}\@gtempa
   \if@nobreak \ifvmode\nobreak\fi\fi\fi\@esphack}
        \gdef\@eqnlabel{#1}}
\def\@eqnlabel{}
\def\@vacuum{}
\def\draftmarginnote#1{\marginpar{\raggedright\scriptsize\tt#1}}

\def\draft{\oddsidemargin -.5truein
        \def\@oddfoot{\sl preliminary draft \hfil
        \rm\thepage\hfil\sl\today\quad\militarytime}
        \let\@evenfoot\@oddfoot \overfullrule 3pt
        \let\label=\draftlabel
        \let\marginnote=\draftmarginnote
   \def\@eqnnum{(\theequation)\rlap{\kern\marginparsep\tt\@eqnlabel}%
\global\let\@eqnlabel\@vacuum}  }

\def\preprint{\twocolumn\sloppy\flushbottom\parindent 2em
        \leftmargini 2em\leftmarginv .5em\leftmarginvi .5em
        \oddsidemargin -.5in    \evensidemargin -.5in
        \columnsep .4in \footheight 0pt
        \textwidth 10.in        \topmargin  -.4in
        \headheight 12pt \topskip .4in
        \textheight 6.9in \footskip 0pt
        \def\@oddhead{\thepage\hfil\addtocounter{page}{1}\thepage}
        \let\@evenhead\@oddhead \def\@oddfoot{} \def\@evenfoot{} }

\def\numberbysection{\@addtoreset{equation}{section}
        \def\theequation{\thesection.\arabic{equation}}}

\def\underline#1{\relax\ifmmode\@@underline#1\else
        $\@@underline{\hbox{#1}}$\relax\fi}

\def\titlepage{\@restonecolfalse\if@twocolumn\@restonecoltrue\onecolumn
     \else \newpage \fi \thispagestyle{empty}\c@page\z@
        \def\thefootnote{\fnsymbol{footnote}} }

\def\endtitlepage{\if@restonecol\twocolumn \else \newpage \fi
        \def\thefootnote{\arabic{footnote}}
        \setcounter{footnote}{0}}  %\c@footnote\z@ }

\catcode`@=12
\relax

\def\figcap{\section*{Figure Captions\markboth
        {FIGURECAPTIONS}{FIGURECAPTIONS}}\list
        {Figure \arabic{enumi}:\hfill}{\settowidth\labelwidth{Figure
999:}
        \leftmargin\labelwidth
        \advance\leftmargin\labelsep\usecounter{enumi}}}
 \relax
\def\tablecap{\section*{Table Captions\markboth
        {TABLECAPTIONS}{TABLECAPTIONS}}\list
        {Table \arabic{enumi}:\hfill}{\settowidth\labelwidth{Table
999:}
        \leftmargin\labelwidth
        \advance\leftmargin\labelsep\usecounter{enumi}}}
 \relax
\def\reflist{\section*{References\markboth
        {REFLIST}{REFLIST}}\list
        {[\arabic{enumi}]\hfill}{\settowidth\labelwidth{[999]}
        \leftmargin\labelwidth
        \advance\leftmargin\labelsep\usecounter{enumi}}}
 \relax
\makeatletter
\newcounter{pubctr}
\def\publist{\@ifnextchar[{\@publist}{\@@publist}}
\def\@publist[#1]{\list
        {[\arabic{pubctr}]\hfill}{\settowidth\labelwidth{[999]}
        \leftmargin\labelwidth
        \advance\leftmargin\labelsep
        \@nmbrlisttrue\def\@listctr{pubctr}
        \setcounter{pubctr}{#1}\addtocounter{pubctr}{-1}}}
\def\@@publist{\list
        {[\arabic{pubctr}]\hfill}{\settowidth\labelwidth{[999]}
        \leftmargin\labelwidth
        \advance\leftmargin\labelsep
        \@nmbrlisttrue\def\@listctr{pubctr}}}
 \relax
\makeatother
\newskip\humongous \humongous=0pt plus 1000pt minus 1000pt

\newif\ifdtup

\relax

\def\be{\begin{equation}}
\def\ee{\end{equation}}
\def\ba{\begin{eqnarray}}
\def\ea{\end{eqnarray}}

\def\del{\partial}

\def\Tr{\mathrm{Tr}}

\def\a{\alpha}

\def\b{\beta}

\def\g{\gamma}
\def\G{\Gamma}
\def\d{\delta}
\def\D{\Delta}
\def\e{\epsilon}

\def\th{\theta}

\def\m{\mu}
\def\n{\nu}

\def\Om{\Omega}
\def\l{\lambda}
\def\L{\Lambda}
\def\s{\sigma}

\def\no{\noindent}
\def\hb{\hfill\break}
\def\qq{\qquad}

\def\IR{\relax{\rm I\kern-.18em R}}

\def \ha {{1\over 2}}

\def \ov {\over}

\def\II{\relax{\rm 1\kern-.35em1}}
\def\IR{\relax{\rm I\kern-.18em R}}
\def\inv{^{\raise.15ex\hbox{${\scriptscriptstyle -}$}\kern-.05em 1}}

\def\ii{\mathrm{i}}
\begin{document}
\renewcommand{\theequation}{\arabic{equation}}
\renewcommand{\theequation}{\thesection.\arabic{equation}}

\newcommand{\beq}{\begin{equation}}
\newcommand{\eeq}[1]{\label{#1}\end{equation}}
\newcommand{\ber}{\begin{eqnarray}}
\newcommand{\eer}[1]{\label{#1}\end{eqnarray}}
\newcommand{\eqn}[1]{(\ref{#1})}
\begin{titlepage}
\begin{center}

\hfill CCNY-HEP-10/5\\
%\hfill arXiv:yymm.nnnn [hep-th]\\

\vskip  0.7in

{\large \bf High spin limits and non-abelian T-duality}

%{\large \bf High spin limits in sigma-model theories and non-abelian T-duality}

\vskip 0.4in

{\bf Alexios P. Polychronakos$^1$}\phantom{x} and\phantom{x}
{\bf Konstadinos Sfetsos}$^{2}$

\vskip 0.15in

${}^1\!$ Physics Department, City College of the CUNY\\
160 Convent Avenue, New York, NY 10031, USA\\
{\footnotesize{\tt alexios@sci.ccny.cuny.edu}}

\vskip .2in

${}^2\!$
Department of Engineering Sciences, University of Patras\\
26110 Patras, GREECE\\
{\footnotesize{\tt sfetsos@upatras.gr}}\\

\end{center}

\vskip .3in

\centerline{\bf Synopsis}

\no The action of the non-abelian T-dual of the WZW model is related
to an appropriate gauged WZW action via a limiting procedure. We
extend this type of equivalence to other $\s$-models with
non-abelian isometries and their non-abelian T-duals, focusing on
Principal Chiral models. We reinforce and refine this equivalence by
arguing that the non-abelian T-duals are the effective backgrounds
describing states of an appropriate parent theory corresponding to
divergently large highest weight representations. The proof involves
carrying out a subtle limiting procedure in the group
representations and relating them to appropriate limits in the
corresponding backgrounds. We illustrate the general method by
providing several non-trivial examples.

\end{titlepage}
\vfill
\newpage
\setcounter{footnote}{0}
\renewcommand{\thefootnote}{\arabic{footnote}}

\renewcommand{\theequation}{\thesection.\arabic{equation}}

%\setlength{\baselineskip}{.7cm} \setlength{\parskip}{.2cm}

%% Section 1:
\setcounter{section}{0}

\def\baselinestretch{1.2}
\baselineskip 20 pt %17.5 pt
\noindent

%%%%%%%%%%%%%%%%%%%%%%%%%%%%%%%%%%%%%%%%%

\tableofcontents

\section{Introduction and conclusions}

An important achievement of string theory is that it can describe spacetime physics
at the quantum level beyond the General Theory of Relativity.
The most appealing class of models admitting an exact string theoretical
description is based on coset $G/H$
conformal field theories (CFTs) \cite{coset} that admit
a spacetime interpretation via the gauged WZW models \cite{gwzwac}.

\no
In physical applications one deals with
field equations. The generic absence, however, of isometries in
the gravitational backgrounds corresponding to $G/H$ coset models
makes them unsolvable with any of the traditional methods.
This deficiency is not a problem in low dimensional coset models,
such as the prototype example of
a two-dimensional black hole in \cite{WittenBH}, or models in which the
subgroup that is being gauged is abelian. It becomes, nevertheless, a major hurdle
when the gauge group is non-abelian (see, for instance, \cite{BaSfe, Lugo}).

\no
In recent work we developed a method that overcomes this problem using
techniques based on the rich, albeit not manifest, underlying group theoretic structure \cite{PolSfe1}.
We gave the general procedure and, in addition, we presented explicit results
for the background corresponding to the $SU(2)_{k_1}\times SU(2)_{k_2}/SU(2)_{k_1+k_2}$ model.

\no
In our present work we focus on the sector of the theory corresponding to representations with
divergently large values of highest weight. This is a consistent
sector and admits a description in terms of
an effective gravitational background, provided that a correlated limit in the levels
is taken so that the eigenenergies of the theory remain finite.
Based on the specific $SU(2)$ example mentioned above,
there are indications \cite{PolSfe1} that these effective gravitational backgrounds are related to
the so-called non-abelian T-duals of the WZW backgrounds.
This is further supported by the fact that the gauged WZW action for the
coset $(G_k \times  H_\ell)/H_{k+\ell}$ is equivalent
in the $\ell\to \infty$ limit, to the action for the non-abelian T-dual of
the WZW model for $G_k$ with respect to the subgroup $H$ \cite{gwzwsfe}.

\no
In the present paper we reinforce this relation by
considering the above limit at the level of the states of the theories.
Specifically, we construct the eigenstates of the scalar equation for the background fields of the coset theory and
carefully take the large spin limit. We demonstrate that these states
solve the scalar wave equation for the effective limiting
background, or, equivalently, for the non-abelian dual of the original WZW model for $G_k$.
We also extend this equivalence to other $\s$-models
with non-abelian isometries and their non-abelian T-duals focusing, in particular,
on Principal Chiral models. In our discussion we present general arguments and give explicit results.

\no
Our results improve our understanding of non-abelian T-duality \cite{nadual, duearl, fratse}
which, unlike the abelian T-duality originating in a string context in \cite{Buscher},
has remained in comparison rather poorly understood in spite of a substantial body of work, e.g.
%\cite{GiRo,Alvarez:1994zr, CurtZach,Lozano:1995jx,Sfetsos:1996pm}.
\cite{GiRo}-\cite{Bossard:2000xq}.
In particular, one may now consider these transformations as
generating effective backgrounds for describing consistent sectors
of some parent theories in the limit of infinite highest weight
representations. In fact, this is the physical reason for the fact
that the non-abelian T-duality transformation is non-invertible at
the level of its path integral formulation.

\section{Gauged WZW models and non-abelian T-duality }
\setcounter{footnote}{0}
\renewcommand{\thefootnote}{\arabic{footnote}}

In this section we briefly review the relation of the gauged WZW
models and the non-abelian duals of WZW models at the level of their classical actions.
%by following \cite{gwzwsfe}.

\no
Consider coset models of the type $(G_k\times H_\ell)/H_{k+\ell}$,
with the subgroup $H$ appropriately embedded into the direct product of the groups $G\times H$.
The gauged WZW action is \cite{gwzwac}
\ba
&& S_{\rm gWZW}(g,h,A_\pm)  = k I_0(g) + \ell I_0(h)
\nonumber\\
&& \phantom{xxxx} + {1\ov \pi} \int_M \Tr \Big[k A_- \del_+ g g^{-1} + \ell A_- \del_+ h h^{-1}
 - k A_+ g^{-1} \del_- g
\label{gwwzw1}\\
&&
\phantom{xxxxxxxx} - \ell A_+ h^{-1} \del_- h
+ k  A_- g A_+ g^{-1} + \ell  A_- h A_+ h^{-1}
- (k+\ell) A_- A_+\Big]\ ,
\nonumber
\ea
where $g$ and $h$ are elements of the groups $G$ and $H$, respectively,
parametrized by a total of $\dim(G) + \dim(H)$ variables $X^M$, and
$I_0(g)$ and $I_0(h)$ are the corresponding WZW actions.
The gauge fields $A_\pm $ also take values in the Lie algebra of $H$ and
the above action is invariant under the gauge transformations
\be
g\to \L^{-1} g \L \ ,\quad  h\to \L^{-1} h \L \ ,\quad A_\pm \to \L^{-1} A_\pm \L - \L^{-1}\del_\pm \L\ ,
\label{ejhg11}
\ee
for a group element $\L(\s^+,\s^-)\in H$.

\no
The procedure of obtaining a $\s$-model from \eqn{gwwzw1}
involves two steps. Due to the gauge invariance we may gauge-fix
$\dim(H)$ parameters in $g$ and $H$, thus reducing the number of parameters to $\dim G$,
thereafter denoted by $X^\m$.
The gauge fields $A_\pm$ can be integrated out via their equations of motion,
yielding a $\s$-model action
with a metric $G_{\m\n}$, an antisymmetric tensor $B_{\m\n}$ and a dilaton field $\Phi$.
One can give general expressions for all these fields, but this will not be needed for our purposes.

%\subsection{Aspects of non-abelian duality in WZW models}

\no
Let us now turn to non-abelian T-duality applied on WZW actions for some group $G$.
To perform such a transformation on a WZW action we start with the action
\ba
&& S_{\rm nonab}(g,v,A_\pm)  =
k I_0(g) + {k\ov \pi} \int_M \Tr \Big[ A_- \del_+ g g^{-1} - A_+ g^{-1} \del_- g
+ A_- g A_+ g^{-1} - A_- A_+\Big]
\nonumber\\
&&\phantom{xxxxxxxxxxxxx} - i {k\ov \pi} \int_M \Tr(v F_{+-})\ ,
\label{gwwzwnon}
\ea
where the field strength for the gauged fields is defined as
\be
F_{+-}= \del_+ A_- - \del_-A_+ - [A_+,A_-]\ .
\ee
The first line in \eqn{gwwzwnon}
is the usual gauged WZW action for a group $G$ with respect to the vector
action of a subgroup $H$. The second line is just a Lagrange multiplier, with the corresponding fields $v$ in
the Lie algebra of $H$, which forces the field strength $F_{+-}$ to vanish.
The above action is invariant under the gauge transformations
\be
g\to \L^{-1} g \L \ ,\quad v \to \L^{-1} v \L\ ,
\quad A_\pm \to \L^{-1} A_\pm \L - \L^{-1}\del_\pm \L\ ,
\quad F_{+-}\to \L^{-1} F_{+-} \L\ ,
\label{ejhgnon}
\ee
again for a group element $\L(\s^+,\s^-)\in H$,
which is similar to \eqn{ejhg11}.

\no
The dual backgrounds are obtained as in the case of the abelian T-duality
\cite{Buscher}. If we first integrate out the Lagrange multipliers $v$, then this forces $F_{+-}=0$ which
means that locally the gauge fields $A_\pm$ can be set to zero, resulting to the standard
WZW action for the
group $G$. Alternatively, we may integrate over the gauge fields (after we partially integrate the
$v F$-term), as they appear non-dynamically, obtaining a different $\s$-model action.
The gauge invariance \eqn{ejhgnon} can be used to gauge fix $\dim (H)$ parameters among the total
of $\dim(G)+\dim(H)$ parameters in $g$ and in $v$. If $H$ is a proper subgroup of $G$, then one
could choose to gauge fix all parameters among those in $g$. If $H=G$, then necessarily some
of the $v$'s are gauged fixed as well. In any case, the maximum
number of $v$'s that can be fixed is $\dim(H)-{\rm rank}(H)$.

\no
It was shown in \cite{gwzwsfe} that in the limit $\ell\to \infty $ the gauged WZW model
action \eqn{gwwzw1} reduces to the action for the non-abelian duality \eqn{gwwzwnon},
provided an
appropriate limiting procedure is followed. We presently review the essential features for our purposes.
Let's rescale the variables parametrizing the group element $h\in H$ with $\ell$ so that in the limit
of large $\ell$ we have the infinitesimal expansion around the identity as
\be
h = I + i \ {k\ov \ell}\  v + {\cal O}\left(1\ov \ell^2\right)\ .
\label{hllim}
\ee
Substituting into \eqn{gwwzw1} we obtain \eqn{gwwzwnon} with the correct Lagrange multiplier term.
We note that the WZW part of the action $I_0(h)$ does not contribute at all in the above limit.
Hence, at the level of the classical action we have the relation
\be
{G_k \times H_\ell \ov H_{k + \ell}}\bigg|_{\ell\to \infty}
 \ =\ \
{\rm dual\ of}\ G_k\ {\rm with\ respect\ to}\ H\ ({\rm vector})\ .
\label{fhkcco}
\ee
Of course, taking this limit in the background that correspond to the action
\eqn{gwwzw1} (after a gauge fixing and elimination of the gauge fields) is
an equivalent procedure as we will demonstrate in the example of section 4.

\no
One way of thinking of the above limiting procedure is that we focus and explore the area
around the identity element of the group. Since this process is classically well defined,
we expect that the non-abelian T-dual background effectively describes a consistent sector of states of the original theory.
Demonstrating this will put the classical equivalence to a firmer quantum mechanical
footing.

\section{Solving the wave equation}

\setcounter{equation}{0}

The $\s$-models corresponding to general coset models are quite complicated and they
lack isometries.
In \cite{PolSfe1} we developed a general systematic method to solve the field equations for the
associated background fields, based on the underlying group theoretical structure.
We focused for concreteness on the scalar field equation,
which in a background with metric $G_{\m\n}$ and dilaton $\Phi$ is of the form
\be
-{1\ov e^{-2\Phi} \sqrt{G}}\ \del_\m e^{-2\Phi} \sqrt{G} G^{\m\n} \del_\n \Psi = E \Psi\ .
\label{lappp}
\ee
To obtain the general solution to this equation one starts with the irreducible representations (irreps)
of $G \times H$, given by direct products $R \times r$. Then the eigenstates of the Laplacian
on the full group manifold are (the matrix indices $\m$,$\n$ in $r$ below not to be confused with the spacetime indices $\m$,$\n$ above)
\be
R_{\alpha \beta} (g) \, r_{\m\n} (h)\ ,
\ee
with eigenvalues
\be
E(R,r) = \frac{C_2 (R)}{k + g_G }+ \frac{C_2(r)}{\ell+ g_H}\ .
\ee
Under the vector $H$-transformation the above states transform in
the representation
\be
( R \times r ) \times ({\bar R}\times {\bar r} )=
( r_1 \oplus r_2 \oplus \cdots ) \otimes ( {\bar r}_1 \oplus {\bar r}_2
\oplus \cdots )\ ,
\ee
where on the right hand side we decomposed $R \times r $ and its conjugate into irreps $r_i$ of $H$.
We get a singlet from all products of the form $r_i \times \bar r_i$.
Denoting by $C_{\alpha \m}^{a} (R,r;r_i )$ the
Clebsch--Gordan coefficient
projecting the state $\alpha$ of $R$ and the state $\m$ of $r$ into the state
$a$ of $r_i$, we construct coset eigenstates as
\be
\psi_{R,r;r_i} (g,h)  = \sum_{a;\alpha,\beta,\mu,\nu}
C_{\alpha \mu}^{a} (R,r;r_i)
C_{\beta \nu}^{a} (R,r;r_i )
R_{\alpha \beta} (g) r_{\mu \nu} (h)\ ,
\label{fhhqo}
\ee
with eigenvalues
\be
E(R,r;r_i) = \frac{C_2(R)}{k+ g_{G}} + \frac{C_2 (r)}{\ell + g_H} -
\frac{C_2 (r_i)}{k +\ell + g_H}\ ,
\label{excva}
\ee
where $g_G$ and $g_H$ are the dual Coxeter numbers for $G$ and $H$.

\no
By construction the eigenfunctions do not depend on $k$ and $\ell$, so we restrict
ourselves to the semiclassical limit in which the dual Coxeter numbers are ignored
and the backgrounds simplify considerably.

\subsection{The limit of infinite highest weight representations}

Consider representations $r$ of the Lie-algebra for $H$ with high values of
the highest weight which we will denote by $j$. Then necessarily the irreps $r_i$ in its tensor
product with $R$ in the Lie algebra of $G$ (of finite highest weight) have also values
for their highest weight of order $j$.
Then for $j\gg 1$ we may write
\be
C_2(r) = a(r) j^2 + b(r) j +  {\cal O}(1)\ ,
\ee
where the highest power of $j$ is dictated by the fact that the Casimir operator is a quadratic
one.
There is a similar expression for $C_2(r_i)$ with coefficients $a(r_i)=a(r)$ and $b(r_i)$,
but with $j$ replaced by $j+n$, where $n$ is finite.
In the infinite $j$ limit, the eigenvalues \eqn{enrjjj} become infinite unless the level $\ell$ becomes
infinite as well in a way proportional to $j$.
Specifically, let in a convenient parametrization
\be
\ell = {k \ov \d}\ j\ ,
\label{lkjlj}
\ee
where $\d$ is a real positive number.
Then from \eqn{excva} we obtain that
\be
E(R,b(r),b(r_i))=\lim_{j\to \infty} E(R,r;r_i) =
 \frac{C_2(R)}{k} +{\ a(r)(\d-2 n) + b(r)-b(r_i)\ov k}\ \d\ .
\label{kjlew}
\ee
Taking the limit in the eigenfunction \eqn{fhhqo} is more delicate since in involves
the limiting behaviour of the Clebsch--Gordan coefficients, as well as of the representations.
The latter actually should be such that they blow up the region around part of the manifold in
a way that the background fields have a well defined limit as well.

\no
In some sense this limiting procedure is similar to the Penrose limit in which
the geometry around a null geodesic is explored, resulting to a plane wave. Indeed,
the Penrose limit within the context of the AdS/CFT
correspondence also involves a restriction to high spin sectors of
the appropriate gauge theories \cite{BMN}.
However,
unlike the Penrose limit that necessarily requires Minkowski signature, in the present case
we may have Euclidean
signature backgrounds as well.
In both cases,
even if the original background geometry is compact, this property is lost in the limit.

\section{Example: Non-abelian dual of the $SU(2)$ WZW model}
%: $SU(2)_{k_1} \times SU(2)_{k_2} / SU(2)_{k_1+k_2}$}

\setcounter{equation}{0}

\no
We will test the above general ideas in the case of the coset
$SU(2)_{k_1} \times SU(2)_{k_2} / SU(2)_{k_1+k_2}$, for which the general gauged WZW action
\eqn{gwwzw1} for direct product groups can be used.

\subsection{The background geometry}

We first review the construction in \cite{PolSfe1}.
We parametrize the associated group elements in the fundamental representation as
\be
g_1 =
\left(
  \begin{array}{cc}
    \a_0 + i \a_3 & \a_2 + i \a_1 \\
    -\a_2+i \a_1 & \a_0 - i \a_3 \\
  \end{array}
\right)\ ,\qq g_2 =
\left(
  \begin{array}{cc}
    \b_0 + i \b_3 & \b_2 + i \b_1 \\
    -\b_2+i \b_1 & \b_0 - i \b_3 \\
  \end{array}
\right)\ ,
\label{e-2-8a}
\ee
where from unitarity
\be
\a_0^2 + \vec{\a}^2 = 1\ , \qq \b_0^2 + \vec{\b}^2 = 1\ .
\label{e-2-8b}
\ee
We also note the following parametrization for a group element $g\in SU(2)$
\be
g = e^{ {\ii \ov 2} (\phi_1-\phi_2) \s_3} e^{ {\ii \ov 2} \th \s_2} e^{ {\ii \ov 2} (\phi_1+\phi_2) \s_3}
=
\left(
  \begin{array}{cc}
    \cos{\th\ov 2} e^{i\phi_1} & \sin{\th\ov 2} e^{-i \phi_2} \\
   -\sin{\th\ov 2} e^{i\phi_2} & \cos{\th\ov 2} e^{-i \phi_1}\\
  \end{array}
\right)\ ,
\label{e-2-1kj}
\ee
which is also the fundamental $j=1/2$ representation and where the Euler angles are
$\phi =\phi_1-\phi_2$ and $\psi = \phi_1+\phi_2$.

\no
We gauge the diagonal $SU(2)$ subgroup of the full $SU(2) \times SU(2)$ group.
Under this, $\vec \a$ and $\vec \b$ transform as vectors.
The background depends only on invariants of these
three-vectors. They can be chosen to be the three combinations
\ba
&& \a = |\vec{\a}|\ ,\qq
\b = |\vec{\b}|\ ,\qq \g = \vec \a\cdot \vec \b\ ,
\nonumber\\
&& 0\leqslant \a,\b\leqslant 1\ ,\qq |\g|\leqslant \a\b\ .
\label{abfg}
\ea
Then, by following the general procedure, we obtain a
$\s$-model with metric
\ba
ds^2 &=& {k_1 + k_2 \ov (1-\a_0^2 ) (1-\b_0^2 ) - \g^2}
\bigl( \D_{\a\a} d\a_0^2 + \D_{\b\b} d\b_0^2 + \D_{\g\g} d\g^2 \nonumber\\
&& + 2 \D_{\a\b} d\a_0 d\b_0 + 2 \D_{\a\g} d\a_0 d\g + 2 \D_{\b\g} d\b_0 d\g \bigr) \ ,
\label{e-2-10q}
\ea
where
\ba
&&\D_{\a\a} = {(1+r)^2 -r(2+r) \b_0^2 \ov r(1+r)^2}\ ,
\qq \D_{\b\b} = {(1+r^{-1})^2 -r^{-1}(2+r^{-1}) \a_0^2\ov r^{-1} (1+r^{-1})^2}\ ,
\nonumber\\
&&\D_{\g\g} = {1 \ov 2 + r +r^{-1}}\ ,
\qq \phantom{xx}
\D_{\a\b} = \g + {\a_0 \b_0 \ov 2+r+r^{-1}} \ ,
\label{e-2-11q}\\
&&\D_{\a\g} = -{ \b_0 \ov (1+r)^2}\ ,
\qq \phantom{xx} \D_{\b\g} = -{ \a_0 \ov (1+r^{-1})^2} \ ,
\nonumber
\ea
with $r = {k_2/k_1}$.
The antisymmetric tensor is zero and the dilaton reads (up to a constant)
\be
e^{-2 \Phi} =   (1-\a_0^2 ) (1-\b_0^2 ) - \g^2 \ .
\label{e-2-12}
\ee
The background is manifestly invariant under the interchange of $\a_0$ and $\b_0$ and a simultaneous inversion
of the parameter $r$. This symmetry interchanges the two $SU(2)$s.

\subsection{Solving the wave equation}

We present here the general solution of the scalar equation by specializing the
general formula \eqn{fhhqo} and discussion to our case.
%\no
%Consider the fundamental representation of $GL(2,\mathbb{R})$
%\be
%R^{1/2}= \left(
%           \begin{array}{cc}
%             a & b \\
%            c & d \\
%           \end{array}
 %        \right) \ .
%\label{rj2}
%\ee
A general representation $R^j$ of $GL(2,\mathbb{R})$ has matrix elements \cite{Vilenkin}
\ba
R^j_{m_1,m_2} (a,b,c,d) =
\sum_k A^{j}_{m_1,m_2,k}\
  a^{j-m_1-k} d^{j+m_2-k} b^k c^{k+m_1-m_2}\ ,
\label{rjka}
\ea
where
\be
A^{j}_{m_1,m_2,k}
 = {\sqrt{(j+m_1)!(j-m_1)! (j+m_2)!(j-m_2)!}\ov k! (j-m_1-k)! (j+m_2-k)! (k+m_1-m_2)!}\ .
\label{rjka1}
\ee
The summation over $k$ extends to all values for which the factorials have non-negative arguments.
For instance, \eqn{rjka} reproduces the fundamental representation
of $GL(2,\mathbb{R})$
\be
R^{1/2}= \left(
           \begin{array}{cc}
             a & b \\
            c & d \\
           \end{array}
         \right) \ .
\label{rj2}
\ee
For the group $SU(2)$ that we are specifically interested, the fundamental representation
$R^{1/2}$ is identified with
\eqn{e-2-1kj}. In addition, $j$ is a half-integer and $m_1,m_2 = -j,-j+1,\dots, j$. Then,
the above sum is finite and the integer $k$ ranges
between the extreme values ${\rm max}(0,m_2-m_1)$ and ${\rm min}(j+m_2,j-m_1)$.
For the $SU(2)$ case it is customary to use the notation $D^j_{m_1,m_2}(\phi,\th,\psi)$
for the irreps,
which are the so-called $D$-functions. Using the parametrization \eqn{e-2-1kj} these are can be expressed as
\be
D^j_{m_1,m_2}(\phi,\th,\psi) = e^{-i(m_1 \phi + m_2 \psi)} d^j_{m_1,m_2}(\th)\ ,
\label{djm1m2}
\ee
where the Wigner's $d$-matrix can be written in terms of the Jacobi polynomials.
%\ba
%&& d^j_{m_1,m_2}(\th) =   \sqrt{ (j+|m_1-m_2|/2+|m_1+m_2|/2)!\
%(j-|m_1-m_2|/2-|m_1+m_2|/2|!\ov (j-|m_1-m_2|/2+|m_1+m_2|/2)! \ (j+|m_1-m_2|/2-|m_1+m_2|/2)!}
%\nonumber\\
%&&%  \phantom{xxxxxxxx}
%\times (-1)^{\ha (|m_1-m_2| + m_1-m_2)}
%\left(\sin{\th\ov 2}\right)^{|m_1-m_2|}
% \left(\cos{\th\ov 2}\right)^{|m_1+m_2|}
% P_{j-|m_1-m_2|/2-|m_1+m_2)/2}^{|m_1-m_2|,|m_1+m_2|}(\cos\th)\ ,
%\label{wigne}
%\ea
Introducing for notational convenience the non-negative integers
\be
m= |m_1-m_2|\ ,\qq n= |m_1+m_2|\ ,
\label{widef1}
\ee
one can prove that
\ba
&& d^j_{m_1,m_2}(\th) =  % (-1)^{\ha (|m| + m)}\
 \sqrt{ (j+m/2+n/2)!\
(j-m/2-n/2)!\ov (j-m/2+n/2)! \ (j+m/2-n/2)!}
\nonumber\\
&&\phantom{xxxxxxxxx}
\times \ \left(\sin{\th\ov 2}\right)^{m}
\left(\cos{\th\ov 2}\right)^{n}
 P_{j-m/2-n/2}^{m,n}(\cos\th)\ ,
\label{wigne}
\ea
where we should also insert a minus sign after the equality if $m_1-m_2$ is an odd positive integer. The
normalization of the Wigner functions is such that \be \int_0^\pi
d\th\ \sin\th\ d^i_{m_1,m_2}(\th) d^j_{m_1,m_2}(\th) = {\d_{i,j} \ov j+1/2}  \ . \ee

\no
Since we have two $SU(2)$ factors we label the corresponding representations $R^{j_1} (g_1)$ and $R^{j_2} (g_2)$,
where $g_1$ and $g_2$ are the fundamental representations parametrized as in \eqn{e-2-8a}.
Then the general state is \cite{PolSfe1}
\ba
&& \Psi^j_{j_1,j_2} = \sum_{m} \sum_{m_2,n_2=-j_2}^{j_2}
C^{j,m}_{j_1,m-m_2,j_2,m_2}\ C^{j,m}_{j_1,m-n_2,j_2,n_2}\  R^{j_1}_{m-m_2,m-n_2}(g_1)\
R^{j_2}_{m_2,n_2}(g_2)\ ,
\nonumber\\
&&  \phantom{xxxxx}
%{\rm where}\
-{\rm min}(j_1-m_2,j_1-n_2,j)\leqslant m \leqslant {\rm min}(j_1+m_2,j_1+n_2,j)\ .
\label{pjhr1}
\ea
where the $C^{j,m}_{j_1,m_1,j_2,m_2}$ are the Clebsch--Gordan coefficients for a state $|j,m\rangle $
in the diagonal $SU(2)_L$ composed from states $|j_1,m_1\rangle |j_2,m_2\rangle$ in $SU(2)_L\times SU(2)_L$.
Similarly, the $C^{j,m}_{j_1,n_1,j_2,n_2}$ are the Clebsch--Gordan coefficients for a state $|j,m\rangle $
in the diagonal $SU(2)_R$ composed from states $|j_1,n_1\rangle |j_2,n_2\rangle$ in $SU(2)_R\times SU(2)_R$.
The sum is formed in such a way that a singlet of the diagonal $SU(2)_L\times SU(2)_R$ is obtained.
The explicit expression for the Clebsch--Gordan coefficients is \cite{RacahII}
\ba
&& C^{j,m}_{j_1,m_1,j-n_1,m-m_1} = \sum_k (-1)^k \left(
2 j +1\ov 2 j+1 + j_1 -n_1 \right)^{1/2}
\nonumber\\
&& \times\
{[(j_1-n_1)!(j_1+n_1)!(j_1+m_1)!(j_1-m_1)!]^{1/2}\ov
k! (j_1-m_1-k)! (n_1+m_1+k)!(j_1-n_1-k)!}
\label{clg}\\
&& \times\
\left((2 j -j_1-n_1)!(j+m-n_1-m_1)! (j-m-n_1+m_1)! (j+m)!(j-m)!
\ov
(2 j + j_1-n_1)! [(j+m-n_1-m_1-k)! (j-m-j_1+m_1+k)!]^2\right)^{1/2}\ ,
\nonumber
\ea
where the summation extends to all values for which the arguments of the factorials are non-negative.
Note that we have introduced the half integer $n_1$ to parametrize the deviation of the spin
$j_2$ from $j$, obeying $|n_1|\leqslant j_1$.

\no
The state \eqn{pjhr1} has an eigenvalue (in the semiclassical regime) equal to
\be
E^j_{j_1,j_2} = {j_1(j_1+1)\ov k_1} + {j_2(j_2+1)\ov k_2} - {j(j+1)\ov k_1+k_2}\ .
\label{enrjjj}
\ee
Given a pair of values for ($j_1,j_2$) there are $2 j_{\rm min}+1$ values for $j$,
where $j_{\rm min}$ is the minimum
of the $j_i$'s.
The above eigenstates are orthogonal for different values of the
triad $(j_1,j_2,j)$ with respect to the measure
\be
e^{-2\Phi} \sqrt{G}\ d\a_0 \wedge d\b_0 \wedge d\g \sim d\a_0 \wedge d\b_0 \wedge d\g\ .
\ee

\subsection{High spin limit and the corresponding effective geometry}

Of particular interest is the large spin behaviour. We assume that one of the spins
becomes large, whereas the other one is kept finite. For instance,
consider
\be
j_1\gg 1\ , \quad  j_2 = {\rm finite} \quad \Longrightarrow \quad j\gg 1\ .
\label{larspin2}
\ee
In this limit, the eigenvalues \eqn{enrjjj} become infinite unless the level $k_1$ becomes
large as well, but in a way proportional to $j$. Specifically, let
\be
j_1= j-n \ ,\qq k_1 = {k_2\ov \d}\\ j\  ,
\label{lkjl}
\ee
where $n$ is a half-integer and $\d$ a positive real parameter.
Then
\be E_{j_2,n,\d}= \lim_{j\to
\infty} E^j_{j_1,j_2} = {j_2(j_2+1)\ov k_2} +  {\d - 2 n\ov k_2}\ \d\ ,
\label{eiginfi}
\ee
in accordance with the general result \eqn{kjlew}.
Taking the level $k_1\to \infty$ has implications for the geometry supporting these
infinite spin states. It is straightforward to show that in order for the background
to have a good limiting behaviour one should focus on a neighborhood of the manifold.

\no
We focus around $\a_0 =1$ and $\g =0$, by performing first the coordinate transformation
\be
\a_0^2 = 1 - r^2 \left[(x_1+\psi )^2 + x_3^2 \right]\ ,
\qq \g = r (x_1+\psi) \cos\psi\ , \qq \b_0 = \sin \psi\ ,
\ee
followed by the limit $r\to 0$. Then the new variables $x_1$ and $x_3$ that we will
use, instead of $\a_0$ and $\g$, become uncompactified.
In this limit we obtain for the metric and dilaton
\ba
ds^2 & = & k_2\left(d\psi^2 + {\cos^2 \psi\ov x_3^2}d x_1^2
+ {\left( x_3 dx_3 + (\sin\psi\cos\psi +  x_1+\psi) dx_1\right)^2\ov x_3^2\cos^2\psi}\right)\ ,
\nonumber\\
e^{-2 \Phi} &  = &  x_3^2 \cos^2 \psi\ .
\label{non1}
\ea
The above
can be considered as the effective background describing the high spin sector of
the original CFT coset model. It is also the non-abelian T-dual of the $SU(2)$ WZW model with respect to $SU(2)$ (after some renaming of variables it becomes identical to eqs. (6.10)
and (6.11) of \cite{GiRo})

\no
Next we would like to explicitly demonstrate that
the general state \eqn{pjhr1} has a well defined large spin limit that simultaneously solves the scalar wave
equation corresponding to the above limit background.
In this respect we first consider the asymptotic behavior of the Clebsch-Gordan coefficients.
Using
Stirling's formula one may prove from \eqn{clg} the following limit \cite{LimitClGo}
\be
\lim_{j\to \infty} C^{j,m}_{j-n,m-m_2,j_2,m_2}
 = d^{j_2}_{m_2,n}(\zeta)\ ,\quad
\lim_{j\to \infty} C^{j,m}_{j-n,m-n_2,j_2,n_2}
 = d^{j_2}_{n_2,n}(\zeta)\ ,
\quad \cos \zeta = {m\ov j}\ ,
\label{dzlim}
\ee
where $d^{j_2}_{m_2,n}$ and $d^{j_2}_{n_2,n}$
are Wigner's $d$-matrix given in \eqn{wigne}.
Note that in this limit we have by assumption that
\be
j_2,m_2,n_2, n = {\rm finite}\ .
\ee
Hence, remarkably,
in the large spin limit the Clebsch--Gordan coefficients do not trivialize but get associated with
an auxiliary $SU(2)$ irrep of spin equal to the smallest of the three spins that
enter in the Clebsch--Gordan coefficients.
In addition,
the high spin limit turns the summation over $m$ in \eqn{pjhr1} into an integration over the
angular variable $0\leqslant \zeta \leqslant \pi$.

\no
Next we determine the expression for the irrep matrix $R^{j-n}_{m-m_2,m-n_2}(g_1)$ that enters in \eqn{pjhr1}.
To do so we take advantage of the freedom to fix the gauge appropriately so as to facilitate
the evaluation of \eqn{rjka}. We chose the gauge
\be
\a_1=\a_2 = \b_2 =0\ ,
\label{gauc}
\ee
so that the remaining entries are
\be
\a_3 = \a\ , \qq \b_3 = {\g\ov \a} \ ,\qq
\b_1=\sqrt{\b^2 - {\g^2\ov \a^2}} \ .
\ee
From \eqn{rjka} and in the limit \eqn{hllim} (where the r\^ole of $h$ is played here by $g_1$)
we have that
\be
\a_3 = { r v_3 }  =r \sqrt{(x_1+\psi)^2 + x_3^2}\quad \Longrightarrow\quad
 v_3 = \sqrt{(x_1+\psi)^2 + x_3^2}\
\ee
and
\be
\b_1 = {x_3 \cos\psi \ov \sqrt{(x_1+\psi)^2 + x_3^2}}\ ,
\qq \b_3 = {(x_1 + \psi)\cos\psi\ov \sqrt{(x_1+\psi)^2 + x_3^2}}\ ,\qq \b_0 =\sin \psi \ .
\ee
%Therefore the entries $b$ and $c$ in \eqn{rj2} introduce a suppressing factor of $1/k_1^{2k + n_2-m_2}$.
%The only term that contributes non-zero to the eigenstate is the one with $k=0$ and $m_2=n_2$.
Due to \eqn{gauc} the entries $b$ and $c$ in \eqn{rj2} are zero. Therefore,
the only non-vanishing contribution comes from the contributes from the term with $k=0$ and $m_2=n_2$.
In addition,
the entries $a$ and $d$ are, to leading order, of the form $a=d^*\simeq 1+i r v_3 $.
Hence, in this limit and for large $j$, we have that
\be
\lim_{j\to \infty} a^{j-m+m_2} d^{j+m-m_2} = e^{-2 i \d v_3\cos \zeta}\ .
\ee
Also note that the coefficients $A^{j-n}_{m-m_2,m-m_2,0}=1$.
Therefore we obtain the finite sum
\be
\Psi_{j_2,n,\d}(x_1,x_3,\psi)=\lim_{j\to \infty}\Psi^j_{j-n,j_2} = \sum_{m_2=-j_2}^{j_2}
 \G_{j_2,m_2,n,\d}(v_3)\ R^{j_2}_{m_2,m_2}(g_2)\ ,
\ee
where
\ba
 \G_{j_2,m_2,n,\d}(v_3) & = &  j \int_0^\pi d\zeta \sin \zeta \left(d^{j_2}_{m_2,n}(\zeta)\right)^2
 e^{-2 i \d  v_3\cos \zeta}
\nonumber\\
&=& j N^{j_2}_{\a\b}
\int_{-1}^1 dx \ (1-x)^{\a} (1+x)^{\b} \left[P^{\a,\b}_{j-\a/2-\b/2}(x)\right]^2
e^{-2 i \d v_3 x}\ ,
\ea
with the definitions
\be
\a=|m_2-n|\ ,\quad  \b=|m_2+n|\ ,\qq N^{j_2}_{\a\b} = {\left(j_2+{\a+\b\ov 2}\right)! \left(j_2-{\a+\b\ov 2}\right)!
\ov \left(j_2+{\a-\b\ov 2}\right)! \left(j_2+{\b-\a\ov 2}\right)! }\ .
\ee
The overall constant $j$ will be subsequently dropped.

\no
We present below some explicit examples:

\no
For $j_2=0$:
\be
\Psi_{0,0,\d} = {\sin 2\d v_3\ov \d v_3}\ .% \qq E_{0,0,\d}={\d^2\ov k_2}\ .
\ee
For $j_2=\ha$:
\be
\Psi_{1/2,\pm 1/2,\d} = \pm { \b_3\ov \d v_3}\ \cos 2 \d v_3 + {2\d \b_0 v_3\mp \b_3 \ov 2 \d^2 v_3^2}\
\sin 2 \d v_3\ .
\\
%&&  E_{1/2,\pm 1/2,\d} ={3+4 \d(\d\mp 1) \ov 4 k_2} \ .
\ee
For $j_2=1$:
\ba
&& \Psi_{1,\pm 1,\d} = {\b_1^2 - 2 \b_3 (\b_3 \mp 2\d \b_0 v_3 )\ov 2 \d^2 v_3^2}\ \cos 2\d v_3
\nonumber\\
&& \phantom{xxxxxx} +
{2 \b_3^2 -\b_1^2 +  \mp 4 \d \b_0 \b_3 v_3 + 4\d^2 (\b_0^2 - \b_3^2) v_3^2\ov 4 \d^2 v_3^3}\ \sin 2\d v_3\ ,
\\
&& \Psi_{1,0,\d} = {2 \b_3^2-\b_1^2 \ov \d^2   v_3^2}\ \cos 2\d v_3 +
{\b_1^2-2 \b_3^2  + 2 \d^2 (1-2 \b_1^2) v_3^2\ov 2 \d^2 v_3^3}\ \sin 2\d v_3\ .
\nonumber\\
%&&  E_{1,\pm 1,\d} ={2+ \d(\d\mp 2)\ov k_2} \  ,\qq E_{1,0,\d}={2+\d^2 \ov k_2}\ .
\nonumber
\ea
We have checked that these are eigenfunctions of (\ref{lappp}) and the corresponding eigenenergies agree with \eqn{eiginfi}.
Obviously the above expressions rapidly become quite complicated as the spin increases and it would have been
difficult to compute, to say the least, without using the correspondence \eqn{fhkcco}.
In the limit $\d\to 0$ the general state is given by the character of the representation of spin $j_2$, as
it was shown in \cite{PolSfe1}, given by
\ba
\Psi_{j_2} & = & \sum_{m=0}^{[j_2]}
{(2j_2+1)!\ov (2 m+1)!\ (2j_2-2 m)!}\
\b_0^{2(j_2-m)} (\b_0^2-1)^{m}
\nonumber\\
& = & 2^{2 j_2} \b_0^{2 j_2} - 2^{2j_2-2} (2 j_2-1) \b_0^{2 j_2-2} + \cdots = U_{2j_2}(\b_0)\ ,
\label{exppl}\\
E_{j_2} & = & {j_2 (j_2+1)\ov k_2}\ ,
\nonumber
\ea
where $U_{2j_2}(\b_0)$ is the Chebyshev polynomials of the
2nd kind.
We have verified that the above eigenfunctions indeed solve the scalar equation \eqn{lappp} with the
indicated eigenvalues.

\section{Non-abelian duality in non-isotropic cases}
\setcounter{equation}{0}

The idea of using a limiting procedure to take advantage of symmetries in order to solve field
equations can be rather straightforwardly
extended to other $\s$-models in which there is a group theoretical structure.

\no
In the rest of this paper we focus our attention on backgrounds in which the isometry group acts with no isotropy.
In particular, consider the Principal Chiral Model (PCM) \cite{pcmm}
for a group $G$. The $\s$-model action is given by
\be
S(g)= -{k\ov \pi} \int_M \Tr(g^{-1}\del_- g g^{-1} \del_+ g)\ .
\label{pcmac}
\ee
This is invariant under the global $G_L\times G_R$ symmetry
\be
g\to \L_L^{-1} g \L_R\ ,\qq (\L_L,\L_R)\in G\ .
\ee
We would like to find the non-abelian dual of this action corresponding to a subgroup $H_L\in G_L$.
We introduce gauge fields $A_\pm$ in the corresponding Lie algebra and add the appropriate Lagrange
multiplier. The corresponding action is
\be
 S_{\rm nonab}(g,v,A_\pm) = -{k\ov \pi}  \int_M \Tr(g^{-1}D_- g g^{-1} D_+ g) + i \Tr(v F_{+-})\ ,
\label{gh11}
\ee
with the covariant derivatives, corresponding to minimal coupling to the gauge fields, given by
\be
\quad D_\pm g  = \del_\pm g -A_\pm g \ .
\label{coovder}
\ee
The action above is invariant under the local "left" symmetry
\be
g\to \l ^{-1} g\ ,\quad v\to \l^{-1} v \l\ ,\quad
\quad A_\pm\to \l^{-1}A_\pm \l - \l ^{-1}\del_\pm \l \ ,\quad \l(\s^+,\s^-)\in H\ ,
\ee
as well as the global "right" (mainly) symmetry $g\to \l'^{-1} g \L_R$, where $\L_R\in G$ and $\l'$ belongs
to the maximal subgroup of $G$ that commutes with the gauge group $H$.

\no
Similarly to the discussion in section 2 we may reproduce \eqn{gh11} via a limiting procedure.
We introduce an independent gauged WZW action for a group $H$ and start with
\ba
&& S_{\rm Hyb}(g,h,A_\pm) =  -{k\ov \pi}  \int_M \Tr(g^{-1}D_- g g^{-1} D_+ g)
\nonumber\\
& &\phantom{x} + \ell I_0(h) + {\ell \ov \pi} \int_M \Tr \Big[ A_- \del_+ h h^{-1}
-  A_+ h^{-1} \del_- h
+ A_- h A_+ h^{-1} - A_- A_+\Big]\ .
\label{SHyb}
\ea
Then, the limit \eqn{hllim} reproduces the Lagrange multiplier term yielding \eqn{gh11}.

\no
The gauge fields in \eqn{gh11} are non-dynamical and, as before, they can be eliminated via their equations
of motion. We also gauge fix $\dim (H)$ of the parameters.
In this paper we are mainly interested in the case with $H=G$. Then we can choose the gauge $g=I$,
completely getting rid of the parameters in $G$ and being left with a $\s$-model solely
for the Lagrange multipliers $v$. The result is
\be
S = {k\ov \pi} \int \del_+ v_a (K^{-1})^{ab} \del_- v_b \ ,\qq K_{ab} = \d_{ab} + f_{ab}\ ,\quad
f_{ab}\equiv f_{ab}{}^c v_c\ .
\label{vKv}\ee
The process of integrating out the gauge fields introduces an extra factor in the path integral measure given by
\be
e^{-2\Phi}= \det (K)\ ,
\ee
which would have been the dilaton factor if there were a stringy origin of the background. Even though
there is no such interpretation here, it is crucial
that $\Phi$ be included in the measure of the scalar wave equation \eqn{lappp}.

\subsection{Asymmetric coset reduction}

To obtain the states for the scalar sector of the above theory we follow a group theoretic
procedure similar to the one used for the non-abelian duals of WZW in section 3.

\no
Our starting point will be a general model of the type $G^{(1)}_{k_1} \times G^{(2)}_{k_2}/H_{k_1+k_2}$,
where the base manifold $G$ is the product of two groups $G^{(1)}$ and $G^{(2)}$ and
the gauged group $H$ is a subgroup of both $G^{(1)}$ and $G^{(2)}$.
The group reduction is done in an asymmetric way. Specifically,
the configuration space is parametrized by the two group elements ($g_1 ,g_2)$ modulo
the identification
\be
( g_1 , g_2 ) \sim ( h g_1 , h g_2 h^{-1} )
\ ,\qq  g_i\in G^{(i)}\ ,\quad i =1,2\ ,\quad h \in H\ .
\ee
Clearly for $G^{(2)} = H$ this reproduces the structure of (\ref{SHyb}) of the previous section.

\no
We start again with the set of eigenstates of the Laplacian on the
base group manifold, which are given by the matrix elements of direct products
of two irreps of the $G^{(i)}$'s, $R^{(1)} \times R^{(2)}$:
\be
R^{(1)}_{\alpha \beta} (g_1) \, R^{(2)}_{\m\n} (g_2)\ ,
\ee
with eigenvalues
\be
E(R) = \frac{C^{(1)}_2 (R)}{k_1 } + \frac{C^{(2)}_2 (R)}{k_2}\ .
\ee
Under the $H$-transformation defining the coset manifold, the above states transform
in the representation
\be
R^{(1)} \times R^{(2)} \times {\bar R}^{(2)}\ ,
\ee
with the indices $\alpha$, $\m$ and $\n$ transforming under the respective group
factor, index $\beta$ remaining free.
The above direct product must be projected to the singlets of $H$.
Decomposing $R^{(1)}$ into irreps $r_{1i}$ and  $R^{(2)}$ into irreps $r_{2j}$ of $H$,
we must reduce irreps of the form
\be
r_{1i} \times r_{2j} \times {\bar r}_{2k}\ ,
\ee
into singlets. This will be possible, in general, only for specific choices of $i,j,k$;
specifically, whenever the decomposition of $r_{1i} \times r_{2j}$ contains $r_{2k}$.
Assuming this to be the case, we denote $C_{\alpha \m}^{a} (R^{(1)},R^{(2)};r_{2j} )$
one of the Clebsch--Gordan coefficients
projecting the state $\alpha$ of $R^{(1)}$ and the state $\m$ of $R^{(2)}$ into the state
$a$ of $r_i$ (there could be many, as the product $R^{(1)} \times R^{(2)}$ may contain
more than one copies of $r_{2i}$). We also denote by $P_{\n}^a (R^{(2)}, r_{2j} )$
the projector that projects the state $\n$
of $R^{(2)}$ onto the state $a$ of $r_{2j}$. We can then construct coset eigenstates as
\be
\psi_{R^{(1)},R^{(2)};r |\beta} (g_1,g_2)  = \sum_{a;\alpha,\mu,\nu}
C_{\alpha \mu}^{a} (R^{(1)},R^{(2)};r)
P_{\nu}^{a} (R^{(2)},r )
R^{(1)}_{\alpha \beta} (g_1) R^{(2)}_{\mu \nu} (g_2)\ .
\label{fhhqo1}
\ee
The above states have two degeneracy indices: $\beta$, which, as we stated, does not
participate in the $H$-transformation and remains free, and an (unmarked) extra index
enumerating the various copies of $r$ contained in $R^{(1)} \times R^{(2)}$.

\no
A special case of interest, analogous to the vector coset case studied in the previous paper,
is the one where $G^{(1)} = G^{(2)} =H \equiv G$. Then the decomposition of $R^{(1)}$ and $R^{(2)}$ into irreps $r_i$ is not
required and we only need the Clebsch--Gordan coefficients
$C_{\alpha \m}^{\n} (R^{(1)},R^{(2)})$
projecting the state $\alpha$ of $R^{(1)}$ and the state $\m$ of $R^{(2)}$ into the state
$\n$ of $R^{(2)}$ (assuming $R^{(1)} \times R^{(2)}$ contains $ R^{(2)}$).
These are the same as the Clebsch--Gordan coefficients
$C_{\m \n}^{\alpha} (R^{(1)},R^{(2)})$
projecting the state $\m$ of $R^{(2)}$ and the state
$\n$ of ${\bar R}^{(2)}$ into the state $\alpha$ of $R^{(1)}$.
The eigenstates are
\be
\psi_{R^{(1)},R^{(2)};\beta} (g_1,g_2)  = \sum_{\alpha,\mu,\nu}
C_{\a\m}^{\n} (R^{(1)},R^{(2)})
R^{(1)}_{\alpha \beta} (g_1) R^{(2)}_{\mu \nu} (g_2)\ .
\label{fhhqo2}
\ee
%The interpretation of the above formula is that out of the
%$d_{R^{(1)}} \times d_{R^{(2)}}$--dimensional multiplet of states for the "left" indices in
%$R^{(1)}_{\a \b} (g_1) R^{(2)}_{\mu \nu} (g_2)$
%the $d_{R^{(2)}}$--dimensional submultiplet transforming in the $R^{(2)}$ representation
%is selected (labeled by the index $\n$).
Further, since $G^{(1)}=H$, we can use the gauge symmetry to completely gauge fix
the first field to $g_1 =\mathbb{I}$ (as was done previously to arrive at (\ref{vKv})) , in which
case $R^{(1)}_{\alpha \beta} (g_1 =\mathbb{I}) = \delta_{\alpha \beta}$. Then, the expression
for the eigenstates simplifies further to
\be
\psi_{R^{(1)},R^{(2)};\b}\big|_{\rm g.f.}  = \sum_{\mu,\nu}
C_{\b\mu}^{\n} (R^{(1)},R^{(2)}) R^{(2)}_{\mu \nu} (g_2)  \ .
\label{fhhqo2b}
\ee
For $R^{(1)} = \mathbb{I}$ (singlet) the summation becomes a trace and the states
become the (conjugation-invariant) characters of $G$.

\no
The identification of irreps $R^{(1)}$ and $R^{(2)}$ such that their product contain $R^{(2)}$
is a matter of group theory for the group $G$. For $G=SU(N)$ an obvious constraint is that
$R^{(1)}$ should have zero $Z_N$ charge (the number of boxes in its Young tableau should
be a multiple of $N$).

\no
As an aside, we note that exactly the same group theory selection rules and eigenstates
arise in the case
of the spin-Calogero--Sutherland model obtaining from matrix models or two-dimensional
Yang--Mills theory, in which case $R^{(1)}$ is the irrep of $SU(N)$ that encodes the spins
of the particles while $R^{(2)}$ is the irrep that generates the various energy eigenstates
\cite{spinCalo}.

\section{Example: Non-abelian dual of the $SU(2)$ PCM}

\setcounter{equation}{0}

In this case we have representation matrices and group structure constants given by
\be
t_a = {\s_a\ov\sqrt{2}}\ ,\qq f_{abs} = \sqrt{2} \e_{abc} \ ,
\ee
where $\s_a$ are the standard Pauli-matrices. Rescaling as $v_a \to  v_a/\sqrt{2}$ we have that
\be
K_{ab}= \d_{\a\b} + \e_{abc} v_c \quad  \Longrightarrow \quad
(K^{-1})^{ab} = {1\ov 1+ v^2}(\d_{ab} + v_a v_b - \e_{abc} v_c)\ .
\ee
Thus we obtain a $\s$-model (essentially the one computed in \cite{duearl, CurtZach})
with metric
\be
ds^2 = (\d_{ab} + v_a v_b){dv_a dv_b \ov 1+ v^2}\
\ee
and antisymmetric tensor
\be
B_{ab} = -{ \e_{abc} v_c\ov  1+ v^2}\ .
\ee
The dilaton factor is simply $e^{-2\Phi}=1+ v^2$.
This $\s$-model has by construction an $SU(2)$ symmetry corresponding to rotations of the
coordinates $v_a$  that can be made
manifest by introducing spherical coordinates in place of the Cartesian ones.
In particular, for the metric we obtain (for notational conformity $v$ is replaced by $r$)
\be
ds^2 = dr^2  + {r^2\ov 1+ r^2} d\Om_2^2\ .
\ee
This is a smooth space, due to the fact that the isometry acts with no isotropy,
interpolating between $\mathbb{R}^3 $ and $\mathbb{R} \times S^2$.

\subsection{Solving the wave equation}

The scalar wave equation \eqn{lappp} for the above background can be solved explicitly. Indeed,
using spherical coordinates and writing $\Psi= \psi(r) Y_{l,m}(\th,\phi)$,
where the $Y_{l,m}$'s are the standard spherical harmonics, we find that the radial function obeys
\be
{d^2\psi\ov dr^2} + {2\ov r} {d\psi\ov dr} + \left(k^2 -{l (l+1)\ov r^2}\right)\psi  = 0 \ ,
\label{dfgp}
\ee
where
\be
E = k^2 + l(l+1)\ ,\qq k\in \mathbb{R}\ .
\ee
This is the spherical Bessel equation with solutions regular at the origin
the corresponding functions $j_l (kr)$. Hence, the full solution is given by
\be
\Psi_{l,m,k}(r,\th,\phi) = j_l (k r)  Y_{l,m}(\th,\phi)\ ,
\label{solnl}
\ee
which constitute a complete set of orthogonal eigenfunctions.\footnote{For non-abelian T-duals
corresponding to PCM for higher than $SU(2)$ groups the eigenfunctions
are not necessarily completely separable.}

\no
We would like to use \eqn{fhhqo2b} in the appropriate limit that we have described
in order to recover solutions of the wave equation corresponding to the non-abelian model and
in particular the one given above.

\no
Putting $R^{(1)}$ = (spin $j_1$) and $R^{(2)}$ = (spin $j$), and using the gauge $g_1 =\mathbb{I}$ as in
\eqn{fhhqo2b}, we have
\be
\Psi_{j,j_1,n_1}\big |_{\rm g.f.} = \sum_{m} C^{j,m}_{j_1,n_1,j,m-n_1} R^{j}_{m-n_1,m}(g_2)\ ,
\ee
The requirement $R^{(2)} \in R^{(1)} \times R^{(2)}$, which in the present $SU(2)$ case means
$j \in j_1 \times j$, imposes the condition $|j-j_1|\leqslant j \leqslant j+j_1$,
or in other words $0\leqslant j_1\leqslant 2 j$. Since we are eventually interested in
the limit of large $j$ but finite $j_1$, this is not a restriction. In addition, $j_1$
 must necessarily be an integer.
The spin index $n_1$ in the above, taking $2j_1 +1$ values, plays the role of the free
(degeneracy) index $\b$ of \eqn{fhhqo2b}.

\no
Then for $g_2$ we have to use the limit \eqn{hllim} and at the same time take the large spin limit.
Taking the limit \eqn{hllim} in the parametrization \eqn{e-2-1kj} of the group element $g_2$,
implies that\footnote{To avoid proliferation of symbols we set the overall constant on the PCM action
\eqn{pcmac} to one. This way it is also not confused with $k$ introduced in \eqn{dfgp}.}
\be
\phi_1\simeq {1\ov 2\ell}\ v_3\ ,\qq \th \simeq {1\ov \ell}\sqrt{v_1^2+v_2^2}\ ,
\qq \phi_2 =-\tan^{-1}{v_1\ov v_2}\ .
\ee
As previously we will need a linear relation similar to \eqn{lkjl}
\be
 \ell= {j\ov \d}\ ,\qq \d\in \mathbb{R^+}\ ,
\label{elldk}
\ee
so that the infinite level is directly related to the infinite spin limit.
In addition, let
\be
m=j s = j \cos\zeta\ ,\qq |s|\leqslant 1\ ,\quad 0\leqslant \zeta\leqslant \pi\ .
\ee

\no
We compute next the $j\to \infty$ limit of $R^j_{m-n_1,m}$ using the expression \eqn{djm1m2} for it.
First we see that the overall  exponential factor becomes
\be
\lim_{j\to \infty}
e^{-i(m-m_1)(\phi_1-\phi_2) + m (\phi_1+\phi_2)} = e^{-i \phi_2 n_1} e^{-i \d s v_3}\ .
\ee
Taking the corresponding limit in $d^j_{m-n_1,m}$, as given by \eqn{wigne}, requires extra care.
First, we note that using Stirling's formula in the form
\be
(n+a)!\simeq \sqrt{2\pi n}\ n^{n+a}\ e^{-n}\ ,\qq {\rm for}\ n\gg 1\ ,\quad a={\rm finite}\ ,
\ee
we have that
\be
\lim_{j\to \infty} {(j+m/2+n/2)!\
(j-m/2-n/2)!\ov (j-m/2+n/2)! \ (j+m/2-n/2)!} = \left(1+|s|\ov 1-|s|\right)^{|n_1|}\ ,
\ee
where, using the definitions \eqn{widef1}, we have made in the left hand side the
replacements $m\to |n_1|$ and $n\to |2m-n_1|\simeq 2|m|$.
The limiting behaviour of the Jacobi polynomials, also appearing in \eqn{wigne}, is found in terms of
Bessel functions. In general one may show that
\be
\lim_{j\to \infty} P_{\d j}^{\a,\g j +\b-\a} \left(\cos{x\ov j}\right)
= \left(2 \d j\ov \sqrt{\d (\d+\g)} x\right)^\a
J_\a\left(\sqrt{\d (\d+\g)} x\right)\ ,
\label{oriojacbe}
\ee
where $\a, \b, \g$ and $\d$ (not to confuse it with $\d$ introduced
in \eqn{elldk}) are finite constants.\footnote{In our case $\a=|n_1|, \g=2|s|,
\d=1-|s|$ and $\b-\a=\mp n_1$, where the upper (lower) sign agrees with the sign of $m$ (equivalently $s$).}
This identity can be proven by first taking the limit in the
differential equation obeyed by the Jacobi polynomials and showing that after changing variables, as
indicated by the above expression, it reduces to the Bessel equation.
The overall normalization constant can
be fixed by examining the behaviour of both sides of it at small values of $x$.\footnote{In a slight
generalization of \eqn{oriojacbe} one replaces the argument of the Jacobi polynomial
$\cos{x\ov j}$ with any function behaving as $\displaystyle 1-x^2/(2j^2)$ for large $j$.}
Assembling everything we find from \eqn{wigne} that
\be
\lim_{j\to \infty} d^j_{m-n_1,m}= J_{|n_1|}\left(\d \sqrt{1-s^2}  \sqrt{v_1^2+v_2^2}\right)\ .
\ee
In addition, using \eqn{dzlim} and the property of the Clebsch--Cordan coefficients under interchanging
the order in the spin pairs in the lower row, we obtain that
\be
\lim_{j\to \infty} C^{j,m}_{j_1,n_1,j,m-n_1} =  (-1)^{j_1} d_{n_1,0}^{j_1}(\zeta)
\sim (1-s^2)^{|n_1|/2} P_{j_1-|n_1|}^{|n_1|,|n_1|}(s)\sim P^{n_1}_{j_1}(s)\ ,
\ee
where we have used \eqn{wigne} and, in the last step, the well known relation between the
Jacobi polynomials and the associated Legendre functions. Altogether, omitting  a constant
overall factor, we obtain
\be
\Psi_{j_1,n_1}(r,\th,\phi)  =  e^{-i n_1\phi_2}
\int_{-1}^1 ds\ e^{-i \d v_3 s} J_{|n_1|}\left(\d \sqrt{1-s^2} \sqrt{v_1^2+v_2^2}\right)
P_{j_1}^{n_1}(s)\ .
\ee
%\ba
%&& \Psi_{j_1,n_1}(r,\th,\phi)  =  e^{-i n_1\phi_2}
%\nonumber\\
%&&\phantom{xx} \times  \int_{-1}^1 ds\ e^{-i \d v_3 s} (1-s^2)^{|n_1|/2}
%J_{|n_1|}\left(\d \sqrt{1-s^2} \sqrt{v_1^2+v_2^2}\right) P_{j_1-|n_1|}^{|n_1|,|n_1|}(s)\ .
%\ea
This should be a solution of the scalar equation \eqn{lappp} and in particular it
should be just \eqn{solnl}.
The comparison should be made using the spherical coordinates ($r,\th,\phi)$,
in place of the $v_i$'s. After also changing integration variable we obtain
\be
\Psi_{j_1,n_1}(r,\th,\phi) = e^{-i n_1\phi}
 \int_0^\pi d\zeta\ \sin\zeta\ e^{-i\d  r  \cos\th \cos\zeta}
J_{|n_1|}\left(\d  r \sin\th \sin\zeta\right) P_{j_1}^{n_1}(\cos\zeta)\ .
\ee
%\ba
%&& \Psi_{j_1,n_1}(r,\th,\phi) = e^{-i n_1\phi}
%\nonumber\\
%&&\phantom{xx} \times
% \int_0^\pi d\zeta\ (\sin\zeta)^{|n_1\!|+1}\
%e^{-i\d  r  \cos\th \cos\zeta}
%J_{|n_1|}\left(\d  r \sin\th \sin\zeta\right) P_{j_1-|n_1|}^{|n_1|,|n_1|}(\cos\zeta)\ ,
%\ea
The last integral has been computed in \cite{intamaz} and is proportional to
$j_{j_1}(\d r) P^{n_1}_{j_1}(\cos\th)$. Taking into account the overall exponential factor we arrive at
\eqn{solnl}, with the obvious identification of the quantum numbers.
In particular, we see that the (integer) spin $j_1$ plays the role of
the angular momentum $l$ in \eqn{solnl}.

\no
This completes the proof of the recovery of the solutions of the model from the large-spin
limit of the appropriate gauged model.

\section{Concluding remarks and future directions}

In this paper we argued that an appropriate high-level/high-highest weight limit in
parent gauged
$\s$-models reproduces the spectrum of appropriate non-abelian T-duals.
We demonstrated this by explicitly working out examples where the involved
group structures are based on $SU(2)$.
We focused on non-abelian T-duals of WZW models and PCM, but
we believe that our findings can be extended
to all $\s$-models with non-abelian isometries and their T-duals. However, we do not have a general proof of that statement.

\no
It will be interesting to explore further the limiting procedure that we have established.
In particular, for the case of non-abelian duals of WZW models it should be possible to
carry it out in full detail at the exact conformal field theory level.

\no
We conclude by pointing out that the connection to matrix models and integrable systems of the
Calogero type, noticed already in the previous paper \cite{PolSfe1}, persists in the cases
studied presently. Specifically, the states and spectrum identified in section 5.1 for the
case $G^{(1)} = G^{(2)} = H$ are the same as those of spin-Sutherland models.
The possibility for further relations between WZW models and generalized integrable models
remains open and worth exploring.

\newpage
\vskip .0 in
\centerline{ \bf Acknowledgments}

\no
K.S. would like to thank the TH-Division at CERN for hospitality and financial support during a
visit in which part of this work was done. The research of A.P. is supported by an NSF grant.

%%%%%%%%%%%%%%%%%%%%%%%%%%%%%%%%%%%%%%%%%%%%%%%%%%%%%%%%%%%%%%%%%%

\end{document}